\documentclass[twocolumn]{aastex631}
\usepackage{float}  % Required for [H] placement specifier
\usepackage{bm}

\shorttitle{Orbital Landscape of Perturbing Planet Solutions}
\shortauthors{Yahalomi \& Kipping}
\graphicspath{{./}{figures/}}
\begin{document}

\title{Mapping the Orbital Landscape of Perturbing Planet Solutions \\ for Single-Planet Systems with TTVs}

\correspondingauthor{Daniel A. Yahalomi}
\email{daniel.yahalomi@columbia.edu}

\author[0000-0003-4755-584X]{Daniel A. Yahalomi} \thanks{LSST-DA DSFP Fellow}
\affiliation{Department of Astronomy, Columbia University, 550 W 120th St., New York NY 10027, USA}

\author[0000-0002-4365-7366]{David Kipping}
\affiliation{Department of Astronomy, Columbia University, 550 W 120th St., New York NY 10027, USA}



\begin{abstract}

There are now thousands of single-planet systems observed to exhibit transit timing variations (TTVs), yet we largely lack any interpretation of the implied masses responsible for these perturbations. Even when assuming these TTVs are driven by perturbing planets, the solution space is notoriously multi-modal with respect to the perturber's orbital period and there exists no standardized procedure to pinpoint these modes, besides from blind brute force numerical efforts. Using $N$-body simulations with \texttt{TTVFast} and focusing on the dominant periodic signal in the TTVs, we chart out the landscape of these modes and provide analytic predictions for their locations and widths, providing the community with a map for the first time: the TTV circus tent diagram. We then introduce an approach for modeling single-planet TTVs in the low-eccentricity regime, by splitting the orbital period space into a number of uniform prior bins over which there aren't these degeneracies. We show how one can define appropriate orbital period priors for the perturbing planet in order to sufficiently sample the complete parameter space. We demonstrate, analytically, how one can explain the numerical simulations using first-order near mean-motion resonance super-periods, the synodic period, and their aliases -- the expected dominant TTV periods in the low-eccentricity regime. Using a Bayesian framework, we then present a method for determining the optimal solution between TTVs induced by a perturbing planet and TTVs induced by a moon.

% If you start each new sentence on a separate line in your text editor, 
% you'll find it easier to edit later on -- you'll be able to comment out entire
% sentences at once, for example. 

\end{abstract}



\section{Introduction}
\label{sec: intro}

Exoplanet transits occur when a planet passes in front of its host star, blocking a portion of the light. If the timing of the transit isn't perfectly periodic, the planet is said to exhibit transit timing variations (TTVs). If physical, these TTVs are caused by gravitational interactions with other masses, typically assumed to be companion planets \citep[e.g.,][]{Dobrovolskis1996a, Dobrovolskis1996b, Miralda-Escude2002, Holman2005, Agol2005, Lithwick2012, Nesvorny2014, Schmitt2014, Deck2015, AgolFabrycky2018}. In this case, TTVs are the results of minor variations (or wobbles) in the planet's orbit. Stellar activity is the biggest cause of spurious TTV signals, as star spots and other stellar activity signals on rotating stars can mimic a TTV of a physical nature \citep[e.g.,][]{Sanchis-Ojeda2011, Mazeh2013, Szabo2013, Oshagh2013, Holczer2015, Mazeh2015, Ioannidis2016, SiegelRogers2022}. 

Planet-planet TTVs have long been recognized as a valuable observational resource for precise characterization of planets down to even Earth-masses \citep{Agol2005, Holman2005} and have been used for some of the most precise characterization of planetary parameters in multi-planet systems \citep[e.g., TRAPPIST-1 with precision equivalent to radial velocity precision of 2.5 cm s$^{-1}$ in][]{Agol2021}. For a thorough introduction to planet-planet TTVs we point the reader to \citet{AgolFabrycky2018} -- but a brief overview follows. 

In short, the amplitude of TTVs is proportional to a planet's own orbital period as planets gravitationally interact on the the orbital timescale. As the acceleration of a body doesn’t depend on its own mass, TTVs of each planet primarily scale with the masses of the \textit{other} bodies in the system. However, higher-order effects---e.g., the way planet~1 perturbs planet~2 can influence how planet~2 then perturbs planet~1 in return---can introduce weak dependencies on a planet’s own mass. In a two planet system, if we define $P_0$ and $P_1$ as the orbital periods of the two planets, $m_0$, $m_1$, and $m_2$ as the masses of the star and two planets, respectively, and $f_{ij}$ (a function of semi-major axis ratio, $\alpha_{ij}=\mathrm{min}(a_i/a_j,\,a_j/a_i)$) describes the perturbations of planet $j$ on planet $i$ and the angular orbital elements of the planets $\theta_{ij} = (\lambda_i, e_i, \omega_i, I_i, \Omega_i, \lambda_j, e_j, \omega_j, I_j, \Omega_j)$, then to lowest order in mass ratio, the $O-C$ TTVs (observed-minus-calculated timing residual; $\delta t_1$ and $\delta t_2$) can be expressed as

\begin{eqnarray}
    \delta t_1 = P_1 \frac{m_2}{m_0} f_{12}(\alpha_{12}, \, \theta_{12}),\\
    \delta t_2 = P_2 \frac{m_1}{m_0} f_{21}(\alpha_{21}, \, \theta_{21}).
\end{eqnarray}

For more on how to evaluate these functions, see \citet{Nesvorny2008, Nesvorny2009, Nesvorny2010, AgolDeck2016, Deck2016}. If there are more than two planets in the system, the mass-ratios of the planets to the star are sufficiently small, and none of the planet-pairs are in a mean-motion resonance (MMR), then the TTVs are well approximated as a linear combination of the perturbations due to each companion planet. In this case, for $N$ planets, the TTVs for planets $i = 1, 2, ..., N$ can be expressed as

\begin{equation}
\delta t_i = P_i \sum_{j\neq i}^{N} \frac{m_j}{m_0} f_{ij}(\alpha_{ij}, \, \theta_{ij}).
\end{equation}

The largest planet-planet TTVs are induced by orbital period changes that occur due to librations of the system about a MMR. The amplitudes of these TTVs can be calculated via conservation of energy \citep{Agol2005, Holman2010}. Close to resonances, changes to both the semi-major axes and eccentricities lead to TTV cycles with a period that depends on the distance from resonance \citep{Steffen2006phd, Lithwick2012}, often called the ``super-period'' of near-MMR planets. The dominant TTV variation is caused by the system to which its resonance is closest to, for which the critical angles are allowed to move slowly and thus effect the builds up. If a planet with period ratio $P_2/P_1$ is within a few percent of the ratio $j/k$ (integers), then the expected TTV super-period is

\begin{equation} \label{eq: super-period}
    P_{\mathrm{TTV}} = P_{\mathrm{sup}} = \frac{1}{|j/P_1 - k/P_2|}.
\end{equation}

The strength of the resonance scales with the planetary eccentricities raised to a power equal to the order of the resonance, where the order is defined as $|j - k|$; thus, first-order resonances exhibit a linear dependence on eccentricity. Therefore, in the low-eccentricity regime, first-order resonances are expected to dominate. Eccentricities in multi-planet systems have been largely found to be small in \textit{Kepler} data, both via the photoeccentric effect \citep{VanEylen2015} and via TTVs \citep{Hadden2014}. Thus, first-order super-periods are likely the driving force of much of the detectable planet-planet TTVs.

Additional, non-resonant, perturbations occur on the timescale of planetary conjunctions, which is when the planets separation is smallest and thus gravitational attraction is strongest. Conjunctions occur with a period equal to the synodic period,

\begin{equation} \label{eq: conjunction period}
    P_\mathrm{TTV} = P_{syn} = \frac{1}{|1/P_\mathrm{1} - 1/P_\mathrm{2}|}.
\end{equation}

Conjunction-induced TTVs (a.k.a. synodic or ``chopping'' TTVs) have smaller amplitudes than near-MMR TTVs as they do not add coherently and commonly show TTVs that alternate early and late, superimposed on the larger amplitude near-MMR TTVs. Conjunction-induced TTVs have been demonstrated to be a useful tool in breaking the mass-eccentricity degeneracy inherent in near-MMR TTVs \citep{Nesvorny2014, Schmitt2014, Deck2015}.

While TTVs are observed as discrete timing deviations at the specific epochs of individual transits, they can be interpreted as samples of an underlying continuous function describing the planet’s longitude deviation. Conceptually, the continuous TTV signal corresponds to the deviation that would be observed by hypothetical observers co-orbiting with the planet, i.e., moving around the star with the same orbital period as the planet itself. Such observers would continuously measure the planet's phase deviations relative to a strictly Keplerian orbit, whereas real observations can only sample these deviations at the discrete transit times.

Due to these sampling limitations of real-world transit observations, aliases are frequent in TTV observations. Transits can only be observed with a minimum period equal to the planet's own orbital period. Thus, the Nyquist period for TTVs, which defines the minimum recoverable period for a dataset, is equal to twice the orbital period of the transiting planet \citep{Nyquist1928, Shannon1949}. Any TTV with a faster period that the Nyquist period won't be observed with its true TTV period, but rather at an \textit{aliased} period, greater than its true period and the Nyquist period limit. In order to determine the observable aliased period, we follow the same derivation as presented in \citet{McClellan1998}, and then adopted in \citet{Dawson2010} for radial velocities and subsequently in \citet{Kipping2021} for TTVs. We find that the observed aliased TTV frequency peaks, $\nu$, in terms of the non-aliased physical TTV period, $P_\mathrm{TTV}$, and the period of the transiting exoplanet, $P_\mathrm{trans}$, occur at

\begin{equation}
    \nu = |\frac{1}{P_\mathrm{TTV}} + m \frac{1} {P_\mathrm{trans}}|,
\end{equation}

where $m$ is a non-zero real integer. Or, in terms of observable aliased TTV periods, $\bar{P}_\mathrm{TTV}$, we have

\begin{equation} \label{eq: alias}
    \bar{P}_\mathrm{TTV} = \frac{1}{\nu} = \frac{1}{|\frac{1}{P_\mathrm{TTV}} + m \frac{1}{P_\mathrm{trans}}|}.
\end{equation}

Observing both planets transit is beneficial in characterizing a near-resonant TTV signal, as the phases of the planetary transits and the TTV signals can be compared and thus ambiguity about the period of the perturbing planet is removed \citep{Lithwick2012}. When only a single planet transits, the measured TTVs could result from a perturbing planet being close to a number of different resonances with the transiting planet \citep{Meschiari2010}. \citet{Nesvorny2012} were the first to successfully discover and completely classify a non-transiting planet via TTVs (and transit duration variations: TDVs) in the Kepler-46 (a.k.a. KOI-872) system. In order to do so, they Fourier-decomposed the TTVs of the transiting planet into at least four significant sinusoids, which all could be identified as the interaction with the non-transiting planet via a different resonance. Finally, TDVs were used to break a degeneracy between two possible near-resonant solutions. 

As discussed in \citet{KippingYahalomi2022} and \citet{Yahalomi2024}, TTVs are frequently observed in single transiting planetary systems, where multiple solutions are possible for the unseen perturbing planet's period via resonant interactions. In this situation, the parameter space is highly degenerate and multi-modal with respect to the unseen planet's orbital period. While nested sampling methods are quite powerful at sampling multi-modal parameter spaces, we've found that they are unable to effectively sample different resonant commensurable periods, when given a wide uninformative prior on the perturbing planet's orbital period. This problem becomes additionally difficult when expanding the realm of possible solutions beyond planet-planet TTVs -- as TTVs can be additionally induced by moons \citep[e.g.,][]{SartorettiSchneider1999, Simon2007, Kipping2009, Kipping2009_ttv_moon_I, Kipping2009_ttv_moon_II, AwiphanKerins2013, Heller2014, Heller2016, KippingTeachey2020}. In order to perform model comparison between planet-planet TTVs and planet-moon TTVs, given an observed TTV signal in a single transiting planet system, it is essential to first sufficiently sample the complete parameter space of both possible solutions.

Here, we aim to classify the orbital landscape of planet-planet TTVs by mapping the solution space for TTV period as a function of orbital period ratios. This allows us to identify the multi-modalities, which can in turn enable modelers to adopt appropriate priors in order to effectively sample the complete parameter space. One could then run multiple samplers, with priors set around the different commensurable near-resonant periods and then compare the resulting \textit{a-posteriori} solutions in order to classify the orbital characteristics of an unseen planetary companion.

In what follows, in Section~\ref{sec: numerical}, we present numerical $N$-body simulations to investigate the TTV period versus orbital period space. In Section~\ref{sec: analytic} we discuss the analytic solutions expected due to near MMR and synodic TTVs in the low-eccentricity regime. In Section~\ref{sec: discussion}, we compare our numerical and analytic results and discuss the dynamical interpretation of the numerical simulations. Finally, in Section~\ref{sec: model comparison} we present a model comparison approach to differentiating between planet induced TTVs and moon induced TTVs.

\section{Analytic TTV Periods}
\label{sec: analytic}

\subsection{External Perturber}

\begin{figure}[h]
    \centering
    % First plot
    \begin{minipage}{0.45\textwidth}
        \centering
        \includegraphics[width=\textwidth]{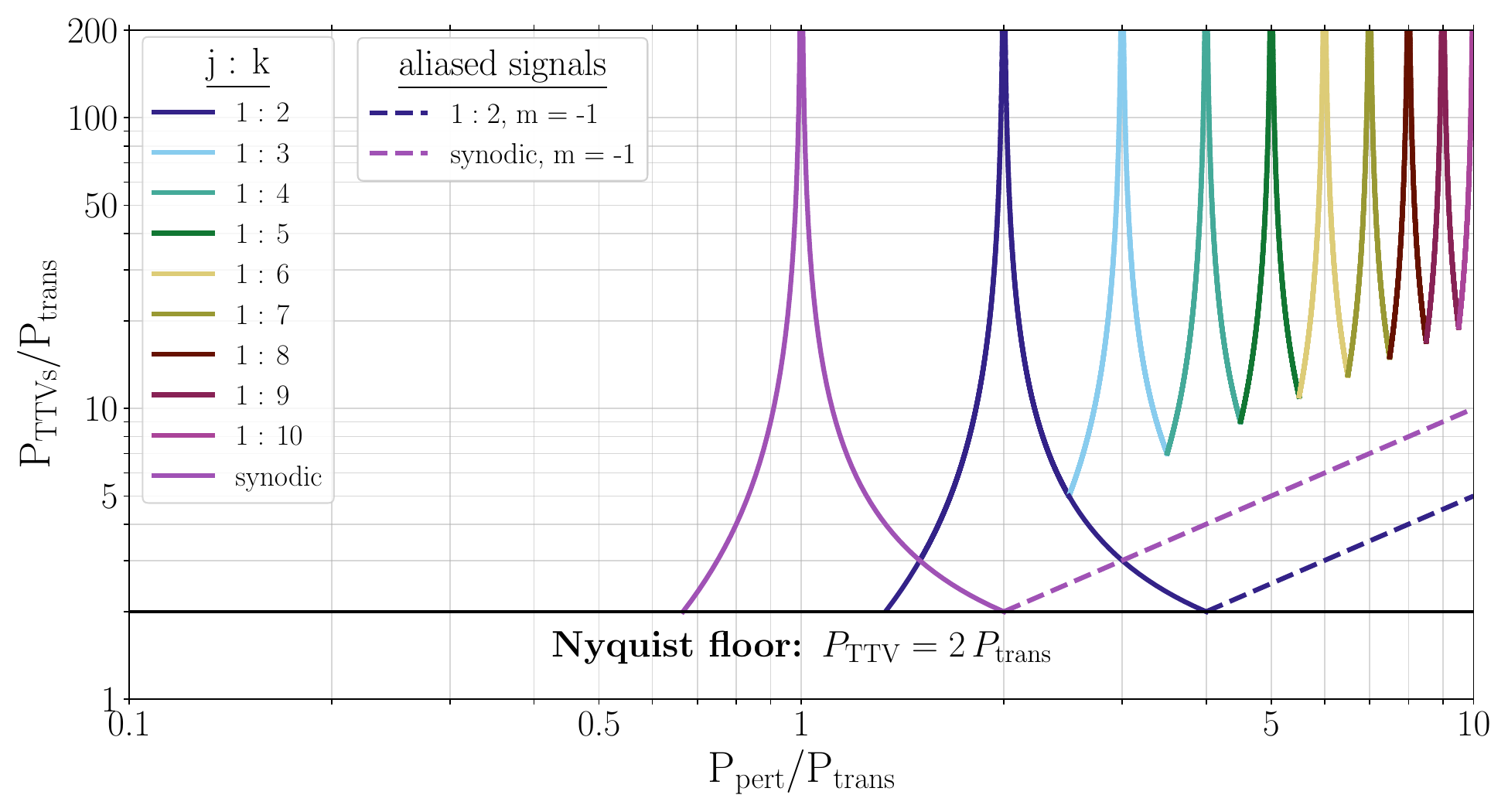}
    \end{minipage}
    \hspace{0.05\textwidth} % Space between plots
    % Second plot
    \begin{minipage}{0.45\textwidth}
        \centering
        \includegraphics[width=\textwidth]{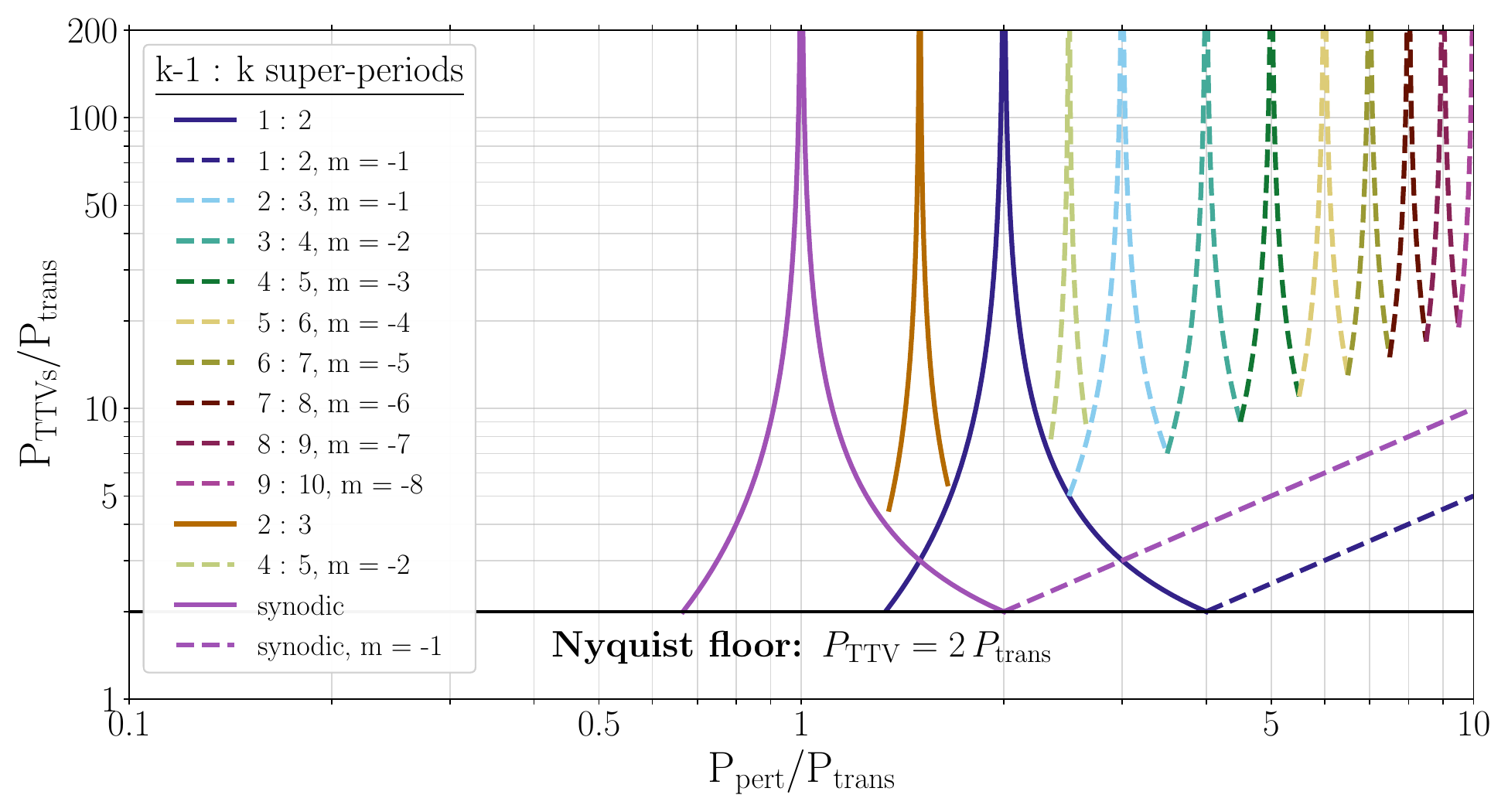}
    \end{minipage}
    
    \caption{Analytic TTVs from super-period equation and their aliases for some $j:k$ near MMR TTVs for external perturbers. Also analytic TTVs from synodic period and its alias. [Top] $1:k$ super-periods shown. [Bottom] Aliases of first-order super-periods shown. Here we show how aliases of first-order ($k-1:k$) super-periods can induce all the same TTV periods as higher order super-periods expected only for large eccentricities.}
    \label{fig: external perturbers}
\end{figure}

Let us start by focusing on an external perturbing planet. We define $P_\mathrm{pert}$ as the period of the perturbing planet and $P_\mathrm{trans}$ as the period of the transiting planet. Since for near MMR TTVs $P_\mathrm{trans}/P_\mathrm{pert} \sim j/k$, we can solve for the nearest $k$ given some $j$ and the period ratio $P_\mathrm{pert}/P_\mathrm{trans}$, via

\begin{equation}
    k = \mathrm{round}(j\,[\, P_\mathrm{pert}/P_\mathrm{trans}]) . 
\end{equation}

Assuming $j$ is 1, we can plot all $1:k$ super-periods for external perturbers with period ratios $1.5 < P_\mathrm{pert}/P_\mathrm{trans} < 10$. This is shown in Figure~\ref{fig: external perturbers}.

As stated earlier, first-order super-periods are expected to dominate near-resonant TTVs in the low-eccentricity regime. Thus, for nearly circular orbits, we wouldn't expect $j:k = 1:k$ for $k>1$ to be frequently detected in TTV data. Thus, we also plot the alias of the $1:2$ in this complete parameter space. Additionally, we plot the synodic TTV and its alias in this complete parameter space. We find that only a solution of $m=-1$ provides an observable solution (i.e., aliased period larger than the Nyquist period), for both the synodic and the $1:2$ MMR in this period ratio regime. Interestingly, we find that the alias of the synodic period is equal to the period of the perturbing planet for period ratios larger than 2 and that the alias of the $1:2$ super-period is equal to half the period of the perturbing planet for period ratios larger than 4. For more discussion on what we dub the ``exoplanet edge,'' see Section~\ref{sec: edge}. These aliases are also shown in Figure~\ref{fig: external perturbers}.

We then investigated other first-order super-periods, and find that for an external perturber with a super-period peak induced by the $j:k$ MMR (where $k>j$), we recover the same super-period peak from the $m = -[k - j - 1]$ alias of the $k-1:k$ MMR. We show this in Figure~\ref{fig: external perturbers}. We also plot the super-period from $j:k = 2:3$ and its $m=-2$ alias.

\subsection{Internal Perturber}

\begin{figure}[h]
    \centering
    % First plot
    \begin{minipage}{0.45\textwidth}
        \centering
        \includegraphics[width=\textwidth]{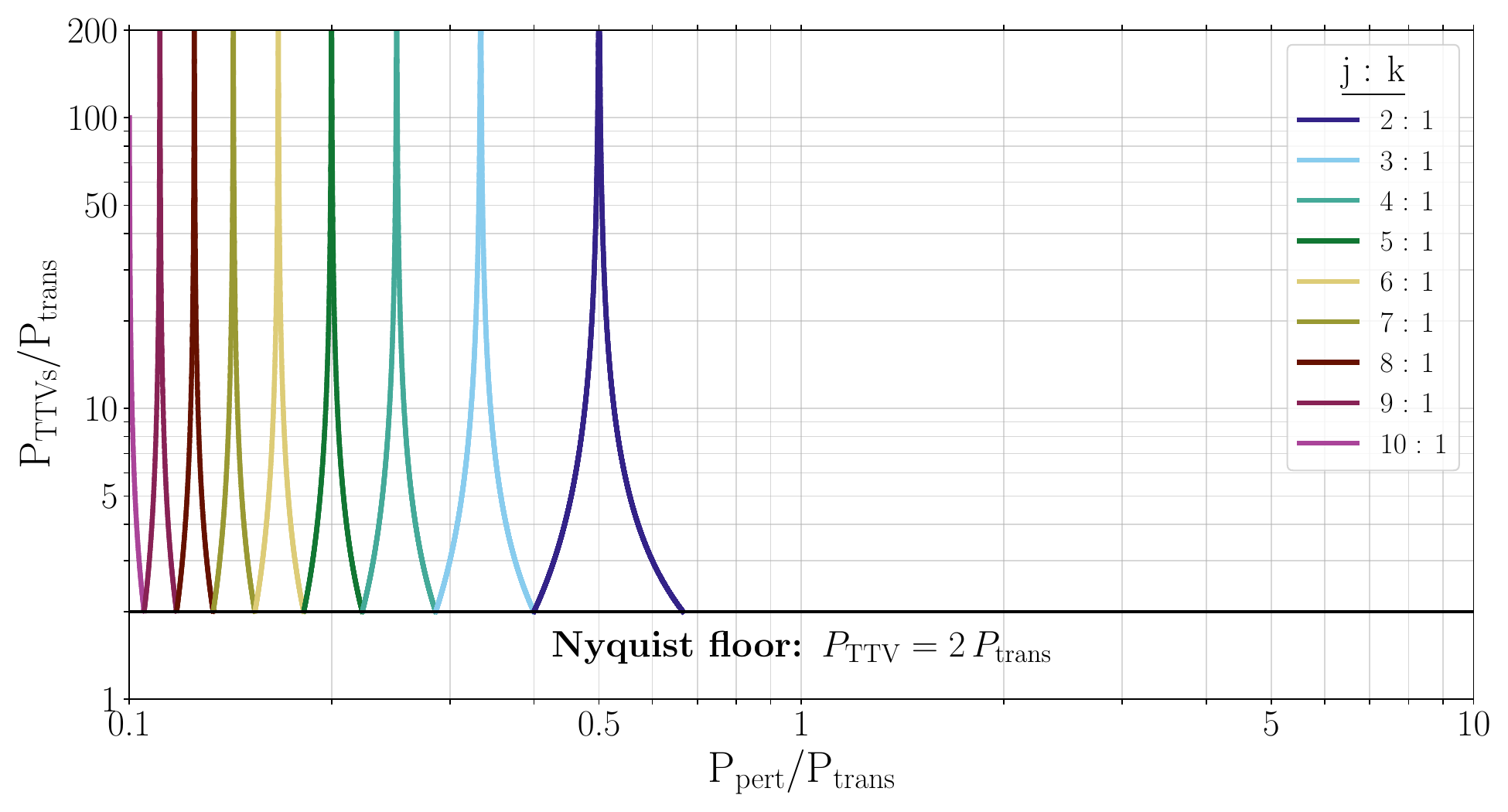}
    \end{minipage}
    \hspace{0.05\textwidth} % Space between plots
    % Second plot
    \begin{minipage}{0.45\textwidth}
        \centering
        \includegraphics[width=\textwidth]{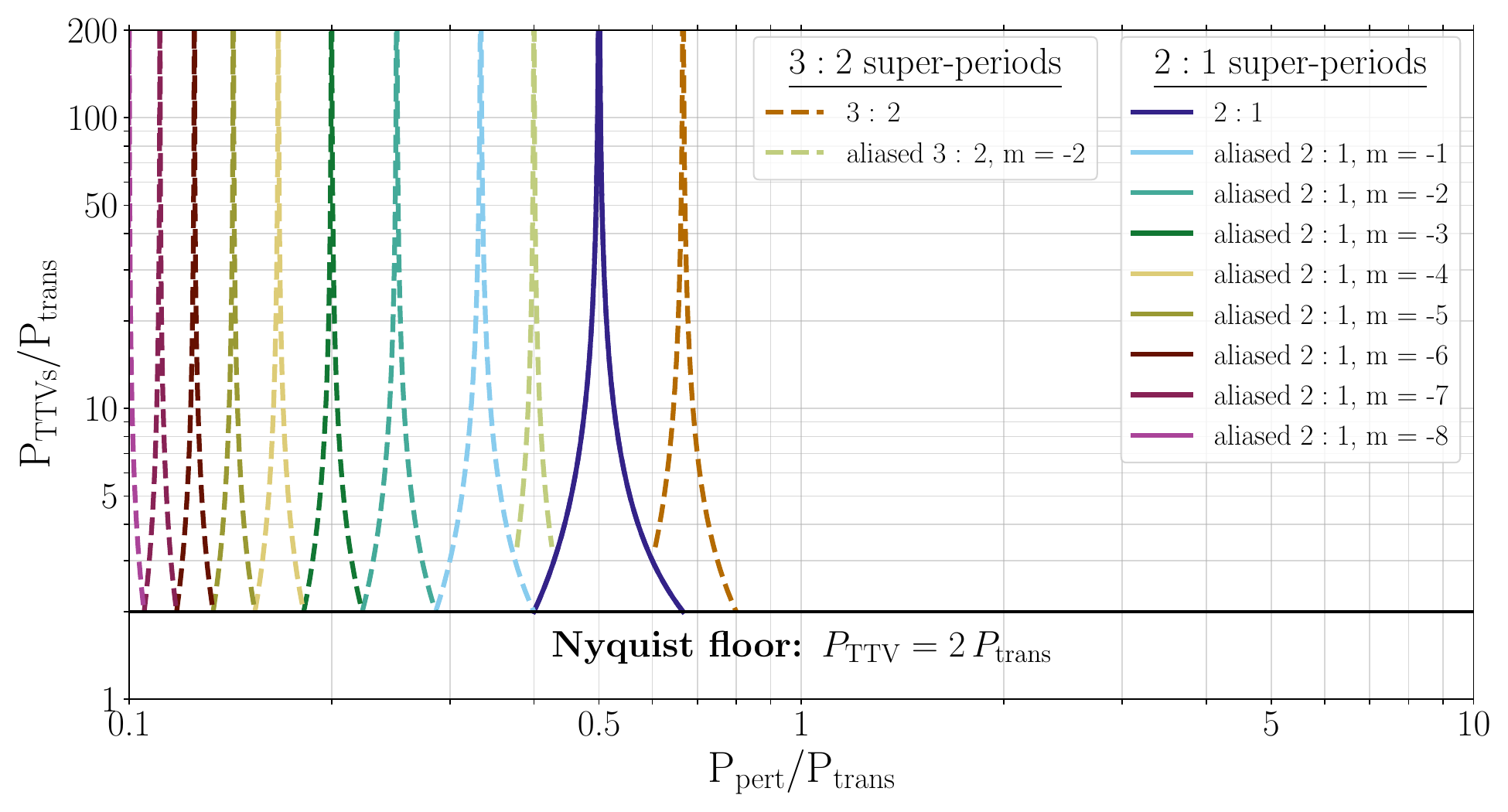}
    \end{minipage}
    
    \caption{Analytic TTVs from super-period equation and their aliases for some $j:k$ near MMR TTVs for internal perturbers. [Top] $1:k$ super-periods shown. [Bottom] Aliases of first-order super-periods shown. Here we show how aliases of first-order ($2:1$) super-periods can induce all the same TTV periods as higher order super-periods expected only for large eccentricities.}
    \label{fig: internal perturbers}
\end{figure}

Now we turn to internal perturbers. Similarly to the external perturber case, we can use the fact that for MMR TTVs $P_\mathrm{trans}/P_\mathrm{pert} \sim j/k$, to solve for the nearest $j$ given some $k$ and the period ratio $P_\mathrm{pert}/P_\mathrm{trans}$, via

\begin{equation}
    j = \mathrm{round}(k / [\, P_\mathrm{pert}/P_\mathrm{trans}]) . 
\end{equation}

Assuming $k$ is 1, we can plot all $j:1$ super-periods for internal perturbers with period ratios $0.1 < P_\mathrm{pert}/P_\mathrm{trans} < 1/1.25$. This is shown in Figure~\ref{fig: internal perturbers}.

Again, in the low-eccentricity regime, we expect first-order resonant TTVs to be the dominant signal -- so let's determine the aliases for the $1:2$ super-period. Similarly to the external perturbers, we find that we can explain the peaks with first-order super-periods and their aliases. Here, we find that the $m=-[j-k-1]$ alias of the $2:1$ super-period are exactly equal to the nearest $j:k$ super-period when orbital period ratios are in the range with period ratios $0.1 < P_\mathrm{pert}/P_\mathrm{trans} < 1/1.25$. This is shown in Figure~\ref{fig: internal perturbers}. We repeat this process for the $3:2$ MMR and plot its resulting super-period and one of its aliases in Figure~\ref{fig: internal perturbers}.

\subsection{Chaotic Region}
As presented in \citet{Deck2013}, first-order resonances become chaotic and thus unstable for close two-planet systems. Specifically, they show that if $\epsilon_p$ is equal to the sum of the masses of the planets in units of the star's mass, then all orbits should be chaotic if their \textit{averaged} period ratio satisfies

\begin{equation}
    P_2/P_1 \lesssim 1 + 2.2 \epsilon_p^{2/7}.
\end{equation}

This criterion is presented as the minimum criterion for widespread chaos -- as this neglects the effect of higher order resonances. Therefore, planets with period ratios internal to this regime, chaotic orbits are expected as overlapping first-order resonances cause changes in the semi-major axes significant enough to oscillate the TTVs between neighboring resonances. Our model comparison approach only works when we can set priors that distinguish near-resonant periods that will not cause these oscillations. Around a Solar mass star, plugging in two Earth mass planets, this corresponds to period ratios of $\sim$1.07 and plugging in two Jupiter mass planets corresponds to period ratios of $\sim$1.4.

\section{Numerical Simulations} \label{sec: numerical}

\begin{figure*}
  \centering
    \includegraphics[width=\textwidth]{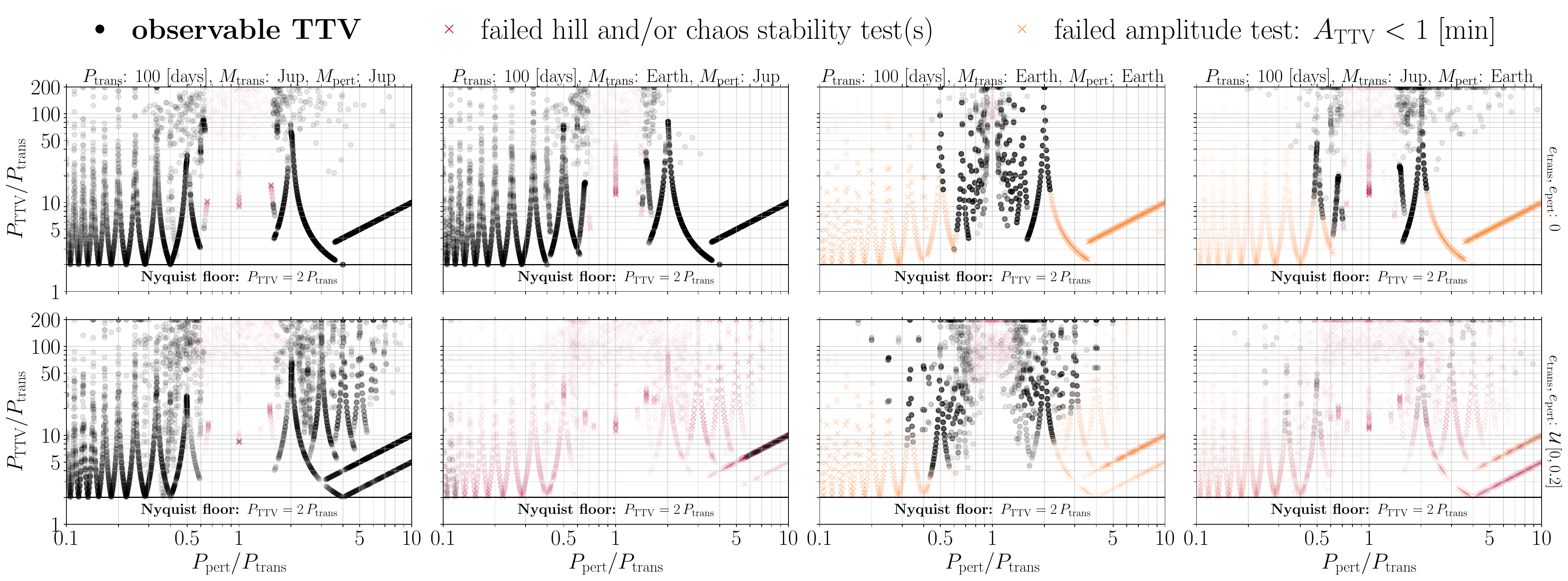}
    \caption{Peak TTV periods recovered via Lomb-Scargle (LS) periodograms fit to transit times simulated with \texttt{TTVFast}. Each column shows a different set of planetary masses. Each row shows a different range of eccentricities. TTVs split into three groups: (i) systems that fail the Hill and/or chaos stability criteria (red x markers), (ii) systems with unreliable TTV amplitudes less than 1 minute (orange x markers), and (iii) those that pass both tests and are thus observable TTVs (blue circle markers).}
    \label{fig: orbital_landscape_mass_sep}
\end{figure*}

\begin{figure*}[htb]
  \centering
    \includegraphics[width=\textwidth]{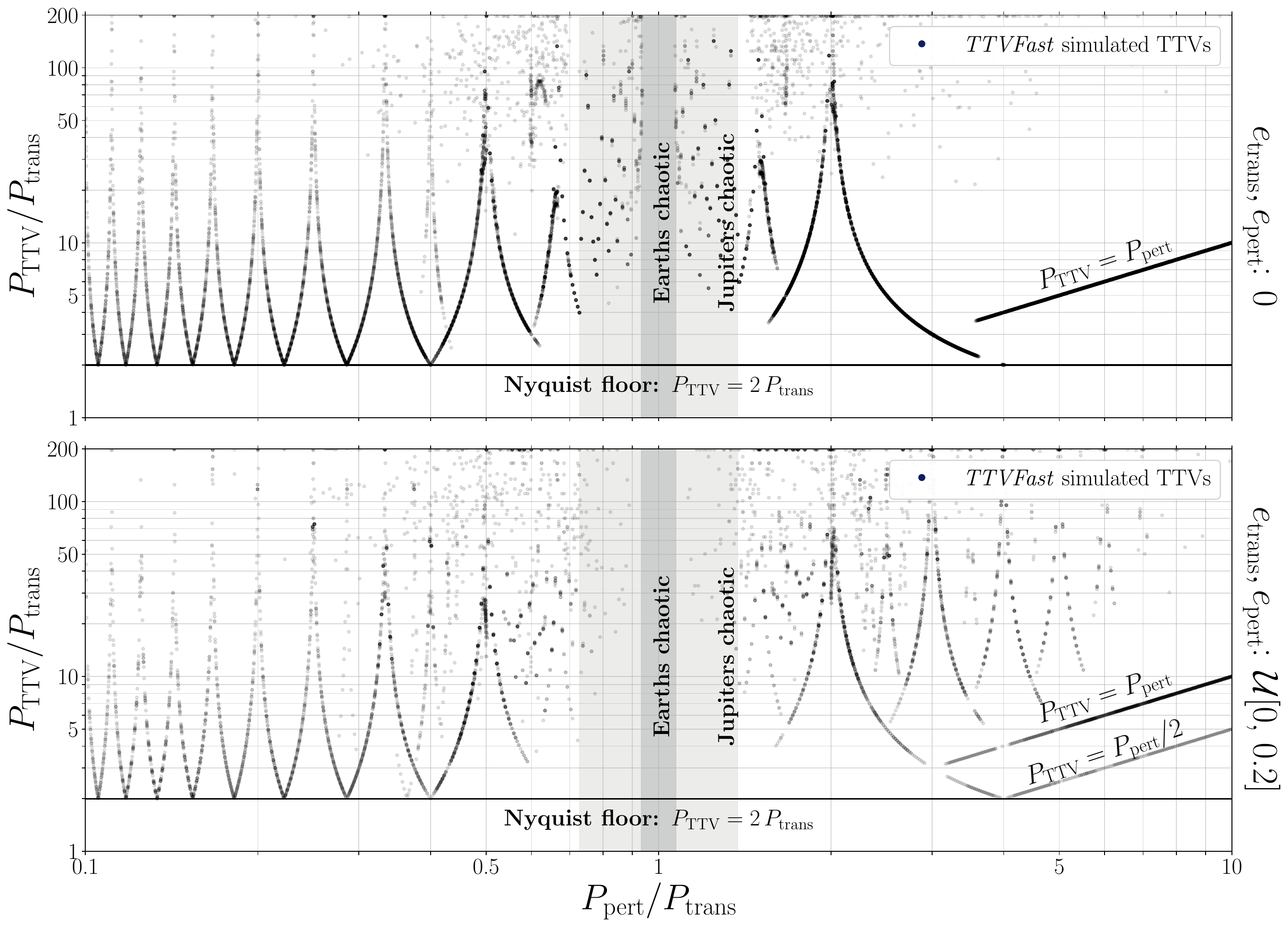}
    \caption{Numerical TTV circus tent diagram. Peak TTV periods recovered via Lomb-Scargle (LS) periodograms fit to \texttt{TTVFast} simulated systems that pass stability and amplitude tests for all combined masses simulated in Figure~\ref{fig: orbital_landscape_mass_sep}. \textbf{[Top:]} \texttt{TTVFast} systems initialized with zero eccentricity \textbf{[Bottom:]} \texttt{TTVFast} systems initialized with eccentricities between 0 and 0.2. We also show the chaotic boundaries as defined by \citet{Deck2013} for two Earth mass and two Jupiter mass planets, respectively, around a Solar mass star.}
    \label{fig: orbital_landscape_combined}
\end{figure*}

We begin by investigating the TTV period versus orbital period ratio space via numerical ($N$-body) simulations to understand the characteristic periods expected for planet induced TTVs. Our first step was to simulate a large number of two-planet systems and map out the period space ($P_{TTV}$, $P_{trans}$, and $P_{pert}$) of the output TTVs using \texttt{TTVFast} \citep{Deck2014}. \texttt{TTVFast} returns a set of transit times for the simulated planets. We then fit a Lomb-Scargle periodogram \citep{Lomb1976, Scargle1982} to these transit times, sampling the periodogram in TTV period space, in order to determine for which sets of orbital parameters would TTVs be observable and the dominant periodic TTV signal in these cases. We describe in detail this process, our simulation assumptions and code, and sampling decisions in what follows.

\subsection{TTVFast}

\texttt{TTVFast} is a dynamical \textit{N}-body simulation package \citep{Deck2014} that given a set of input planet parameters computes a set of predicted transit times. \texttt{TTVFast} uses a symplectic integrator with a Keplerian interpolator for the calculation of transit times \citep{Nesvorny2013}. \texttt{TTVFast} optimizes for computational speed, based on the goal of achieving $\sim$1-10 second precision of transit times. This precision is sufficient for our modeling purposes, as it is less than the typical measurement uncertainty of transit timing surveys \citep{Deck2014}.

As we are interested in the low-eccentricity regime, we simulate two different populations of two-planet systems: (i) zero initial eccentricities for both planets and (ii) small initial eccentricities (e $<$ 0.2). We have done so to take a pragmatic approach, as allowing larger eccentricities increases the dependency on higher order super-periods and thus significantly complicates the multi-modality \citep{MurrayDermott} and as previously explained, demographically, astronomers have found the eccentricities of planets in multi-planet systems are predominantly small. We note that the eccentricity value here is, as is assumed in \texttt{TTVFast}, an instantaneous eccentricity that is a combination of the free and the forced eccentricity values \citep{Deck2014}.

In our simulations, we simulate coplanar orbits. We adopt this assumption as small mutual inclinations tend to have negligible effects on TTVs \citep{Nesvorny2014, AgolDeck2016, HaddenLithwick2016} and known multi-planet systems appear to be mostly coplanar with typical mutual inclinations less than several degrees \citep{Figueira2012, TremaineDong2012, Fabrycky2014}. For all simulations we simulated a Solar-mass star -- but note that the TTV period is scale free with respect to the stellar mass. 

We simulated the perturbing planet with a log$_{10}$-linear grid with 1,000 orbital period values ranging from 1/10 the orbital period of the transiting planet to 100 times the orbital period of the transiting planet. We chose a random initial argument of periastron and mean anomaly for both planets (i.e., random values from the uniform distribution $\mathcal{U}$[0, 360] [deg]). For every simulation, we assume a longitude of ascending node equal to 0 and an inclination equal to 90 [deg].

We fix the orbital period of the transiting planet to 100 days and sample planetary mass pairs of [$M_\mathrm{trans}$, $M_\mathrm{pert}$] equal to: (i) [Jupiter, Jupiter], (ii) [Earth, Jupiter], (iii) [Earth, Earth], and (iv) [Jupiter, Earth]. Modifying the period of the transiting planet will have an impact on the amplitude of the TTV signal, as previously discussed, but it will not impact the TTV period.

Our adopted mass pairs and eccentricity distributions are not intended to reproduce the detailed demographics of known exoplanetary systems, but rather are chosen to span a wide dynamical range relevant for planet–planet TTV interactions. The mass combinations bracket the regimes in which either the transiting planet or the perturber dominates the dynamical forcing, while also including extreme mass ratios that challenge stability. Similarly, the two eccentricity populations (strictly circular and a uniform distribution up to $e=0.2$) are selected not as demographic analogues, but as dynamical probes: circular systems minimize contributions from higher-order harmonics, whereas the $e<0.2$ set introduces modest free and forced eccentricities providing a probe into the higher order eccentricity regime. These choices, therefore, allow us to test the robustness of the TTV period–ratio mapping across a broad dynamical landscape. Importantly, our core results -- specifically, the emergence and location of distinct period–ratio bins and their associated TTV super-periods -- remain largely unchanged when varying the underlying mass and eccentricity distributions. By testing this framework against a wide range of planetary masses, mass ratios, and eccentricities we can thus ensure that this map and modeling framework is robust for a wide range of planetary architectures.

We ran these simulations with \texttt{TTVFast} for a duration equal to 100 times the initial orbital period of the transiting planet with a timestep (dt) equal to 1/20$^\mathrm{th}$ the orbital period of the inner planet as suggested in \citet{Deck2014} as larger time steps can lead to step-size chaos and inaccurate orbits \citep{WisdomHolman1992}. We selected 10 different orbital parameter configurations (i.e., sets of eccentricities in each of the relevant ranges, arguments of periastron, and mean anomalies from the distributions described above). For each of the 2 eccentricity ranges, we ran 10,000 \texttt{TTVFast} simulations (1,000 period ratios and 10 sets of random orbital parameters). In total, we ran 20,000 \texttt{TTVFast} simulations. 

We only keep \texttt{TTVFast} simulations where at least 50 epochs of the transiting planet are returned. This is a conservative cut, as systems in which the inner planet would not have transited at least 50 times during the simulation time scale (100 transit epochs) suggests an orbital period that changed by a factor of greater than two. It is very unlikely, given the long timescale of planetary evolution, that one would observe a stable planet undergoing such rapid dynamical change -- and thus we don't want it to bias our results.

\subsection{Lomb-Scargle Periodogram}
\label{sec: ls}

Following \citet{VanderPlas2018}, we fit a Lomb-Scargle (LS) periodogram \citep{Lomb1976, Scargle1982} to the transit times output from \texttt{TTVFast}. Specifically, we use \texttt{numpy.linalg.solve} to solve the TTV model equation $F(x, \tau, P, \alpha_\mathrm{TTV}, \beta_\mathrm{TTV}, P_\mathrm{TTV})$. If $x$ is the epoch number, $\tau$ is the time of transit minimum, P is the linear ephemeris transit period, $\alpha_\mathrm{TTV}$ \& $\beta_\mathrm{TTV}$ are the amplitude factors (such that the amplitude of the TTV signal is equal to $\alpha_\mathrm{TTV}$ \& $\beta_\mathrm{TTV}$ added in quadrature), and $P_\mathrm{TTV}$ is the dimensionless period of the TTV signal (units of transit epochs), then the linear equation $F(x, \tau, P, \alpha_\mathrm{TTV}, \beta_\mathrm{TTV}, P_\mathrm{TTV})$ used in our LS periodogram is

\begin{eqnarray} \label{eq: TTV_equation}
    \lefteqn{F(x, \tau, P, \alpha_\mathrm{TTV}, \beta_\mathrm{TTV}, P_\mathrm{TTV}) =}  \\
    && \tau + P\,x + \alpha_\mathrm{TTV}\sin \Bigg(\frac{2\pi x}{P_\mathrm{TTV}}\Bigg) + \beta_\mathrm{TTV}\cos \Bigg(\frac{2\pi x}{P_\mathrm{TTV}}\Bigg). \nonumber
\end{eqnarray}

Here, $F(x, \tau, P, \alpha_\mathrm{TTV}, \beta_\mathrm{TTV}, P_\mathrm{TTV})$ becomes a linear equation once we define $P_\mathrm{TTV}$ over a grid with a range between 2 orbits of the transiting planet to twice the number of periods simulated. We setup a TTV period grid evenly spaced in TTV frequency space with a size equal to 10 times the number of transit epochs simulated.

We also use \texttt{numpy.linalg.solve} to solve the null model, $F(x, \tau, P)$ assuming a linear ephemeris (i.e., no TTVs) for the transit times. Here $F(x, \tau, P)$ is

\begin{equation} \label{eq: linear_ephemeris}
    F(x, \tau, P) = \tau + P\,x.
\end{equation}

For each frequency value in the frequency grid, we determine the $\Delta\chi^2$ value by taking the difference between the $\chi^2$ values of the TTV model and the linear ephemeris model. Here, we just assume some fiducial homoscedastic error on the times to evaluate the $\chi^2$ values. As we are always comparing the $\chi^2$ values in determining $\Delta\chi^2$, this assumed uncertainty won't impact the peak solution determination. We then pick the highest $\chi^2$ value in the grid and label this the peak TTV solution. As there are multiple periodic components in TTV data, this doesn't necessarily identify the TTV signal with the peak amplitude. The LS periodogram might pick up a strong signal with a lower amplitude if it has a clearer periodicity or less interference from other signals. Based on existing work on planet-planet TTVs (e.g., see \citet{AgolFabrycky2018} for a summary), it is not unreasonable to expect a dominant periodic component with larger amplitude and then additional harmonic components with smaller amplitudes. Given this, we expect the LS is much more likely to recover this dominant TTV signal with the largest amplitude. Future work should investigate the effect of fitting for multiple periodic signals in the TTVs. Via the LS periodogram, we determine the peak $\Delta\chi^2$ model and save the TTV period and TTV amplitude for each two-planet model simulated with \texttt{TTVFast}. 

\subsection{Removing Non-Observable Systems}

We opted to remove systems that would not be stable over long timescales, as these systems would be unlikely to be observed. We determined stability via two criteria: Hill stability and chaos stability.

To determine the Hill stability criterion, we follow \citet{Petit2018}, using a coplanar approximation. We remove all systems where C$_\mathrm{sys}$ is greater than C$_\mathrm{crit}$. C$_\mathrm{sys}$ is the relative angular momentum deficit (AMD), which quantifies the system’s orbital excitation by combining the eccentricity and inclination contributions of each planet and is defined by Equation~23 of \citet{Petit2018}. C$_\mathrm{crit}$ is the Hill stability criteria in the planetary case, defined by Equation~26 of \citet{Petit2018}. This stability is a function of the mass ratios between the two planets and host star, the periods of the two planets, and the eccentricities of the two planets. 

We also exclude chaotic multi-planet systems based on constraints discussed in \citet{Tamayo2021} and originally presented in \citet{HaddenLithwick2018}. Specifically, we remove all systems where Z$_\mathrm{sys}$ is greater than Z$_\mathrm{crit}$. Z$_\mathrm{sys}$ is the difference in complex eccentricities as defined in Equation~16 of \citet{HaddenLithwick2018}. Z$_\mathrm{crit}$ is the critical Z$_\mathrm{sys}$ value, which provides an approximate formula for the onset of chaos -- as defined in Equation~19 of \citet{HaddenLithwick2018}. This stability is a function of the mass ratios between the two planets and host star, the periods of the two planets, the eccentricities of the two planets, and the longitudes of perihelion of the two planets.

For most of the \textit{Kepler} targets in TTV catalogs, the timing uncertainties are on the order of several minutes or larger \citep{Mazeh2013} and thus we remove all numerically simulated TTVs with amplitudes less than 1 minute. If one wanted to be less conservative, and expand these simulations to higher precision transit timing observations, they could remove systems with amplitudes down to 10 seconds, but with our numerical simulations we would not advise trusting TTV amplitudes any less than this as \texttt{TTVFast} provides transit times with typical scatters of $\sim$1-10 seconds.

We additionally test whether our surviving systems are in steady-states by comparing what fraction of the numerically simulated systems (that pass both stability tests) have best-fit orbital periods within 1\% of the input orbital period. We find that this is largely true, with $\sim$93\% of systems passing this $<$1\% orbital period change threshold.

\begin{figure*}[htb]
  \centering
    \includegraphics[width=\textwidth]{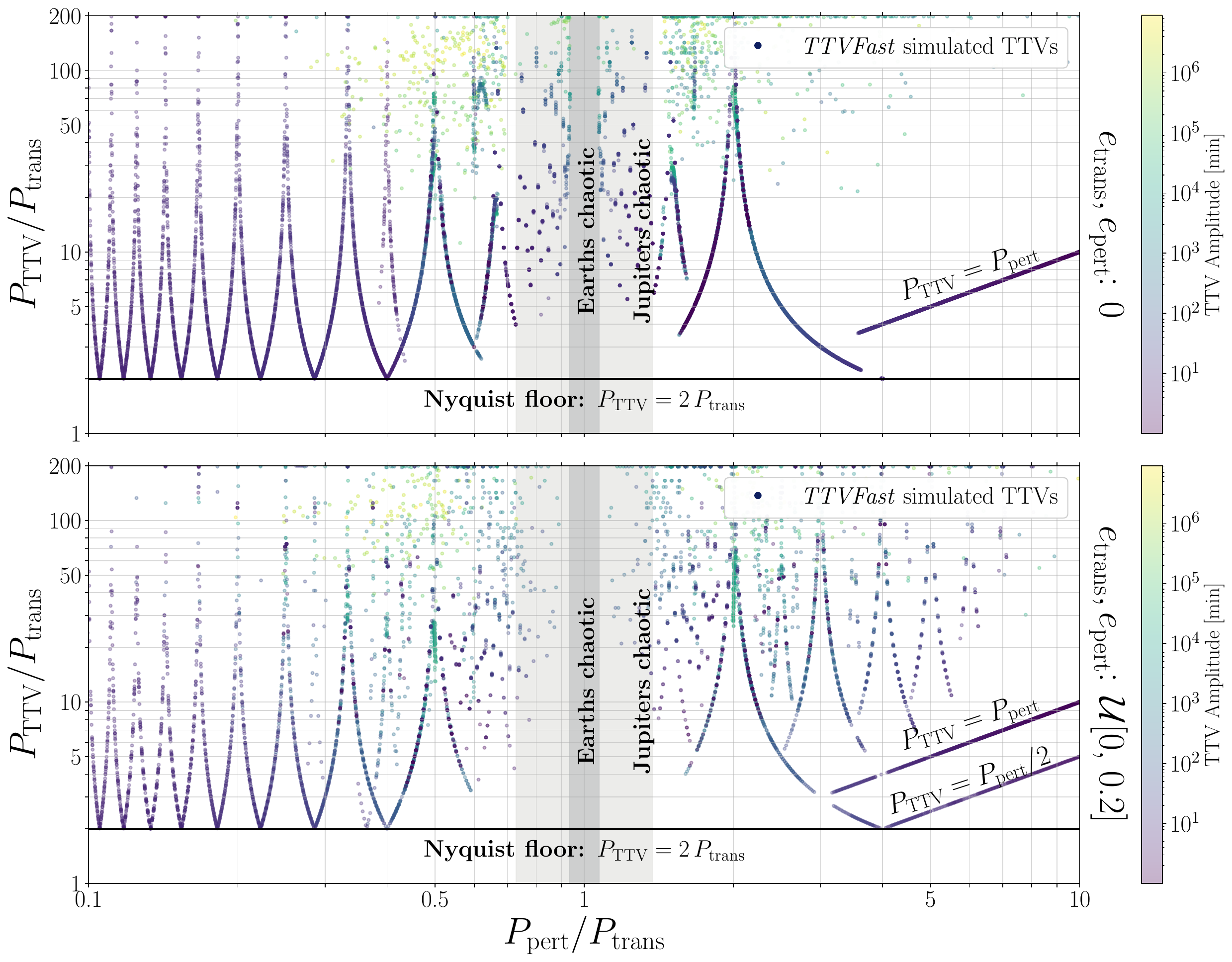}
    \caption{Amplitude sensitive numerical TTV circus tent diagram. Same as Figure~\ref{fig: orbital_landscape_combined}, but with an added colorbar that shows the amplitude of the peak TTV signal identified via the Lomb-Scargle periodogram.}
    \label{fig: orbital_landscape_combined_amplitude}
\end{figure*}

The process of down-selection to observable TTV systems can be seen in Figure~\ref{fig: orbital_landscape_mass_sep}, where we plot TTV period versus perturbing period both normalized by the transiting period. We have separated the data in columns by the planetary masses and in rows by the initialized instantaneous eccentricity ranges initialized for the two planets. Additionally, in Figure~\ref{fig: orbital_landscape_combined}, we show the combined results for all surviving TTVs, separated by the two eccentricity bins used. Finally, Figure~\ref{fig: orbital_landscape_combined_amplitude} shows the same results shown in Figure~\ref{fig: orbital_landscape_combined}, but now each individual numerical simulation is colored based on the amplitude of the peak TTV signal from the LS fit. We call these diagrams ``TTV circus tent diagrams'' as the peaks produced by super-periods, that cause the period degeneracy, resemble circus tents.

These numerical simulations provide a set of peaks and valleys in TTV period versus orbital period ratio space. These peaks and valleys correspond to individual modes in the solution space. As previously described, planet-planet TTVs are mostly induced by near mean-motion resonance (MMR) observed with a period equal to the super-period. We now aim to explain the observed peaks in the numerical simulations with analytic expressions.

\section{Discussion} \label{sec: discussion}

\subsection{Combining the Analytic and Numerical Solutions}

\begin{figure*}[htb]
  \centering
    \includegraphics[width=\textwidth]{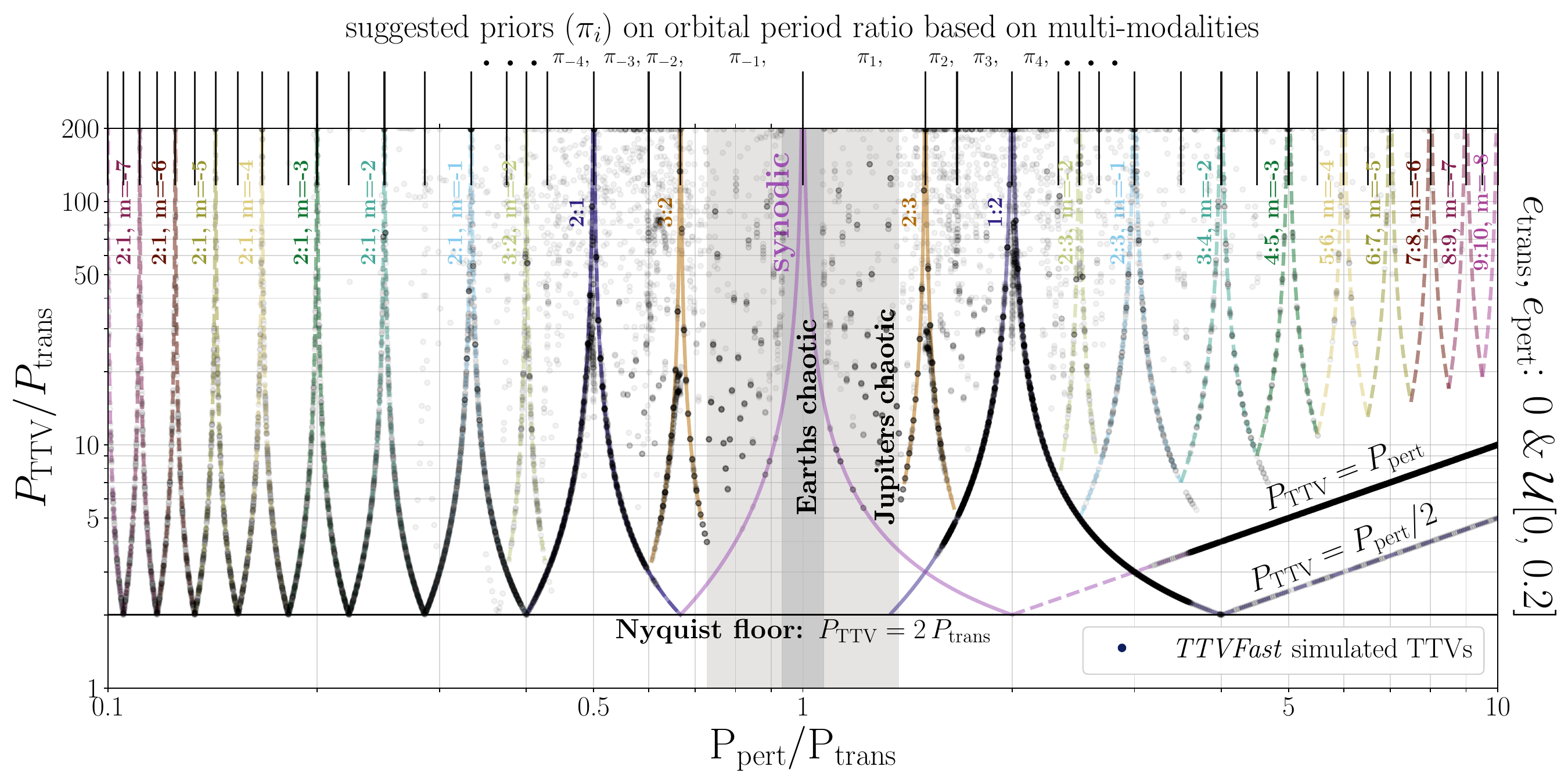}
    \caption{Analytic and numerical TTV circus tent diagram. The analytic model, explained solely by first-order near mean-motion resonant super-periods and the synodic period (and their aliases), fits the numerical results very well. Numerical simulations combine all observable TTVs that passed tests explained in Figure~\ref{fig: orbital_landscape_mass_sep}. We also show the chaotic boundaries as defined by \citet{Deck2013} for two Earth mass and two Jupiter mass planets, respectively, around a Solar mass star. Also recommended orbital period ratio prior boundaries for efficient and accurate modeling.}
    \label{fig: analytic numerical landscape}
\end{figure*}

We can now bring everything together -- incorporating the effects of both internal and external perturbers and mapping the chaotic region as defined by \citet{Deck2013}. We can additionally include the numerical simulations to investigate how well our analytic expression describes the numerical simulations. This can be seen in Figure~\ref{fig: analytic numerical landscape} and the strength of the fit is apparent. This suggests that unsurprisingly, the dominant signal in nearly circular planet-planet TTVs are well described by the analytic first-order MMR super-periods (and their aliases) explained in Section~\ref{sec: analytic}.

There are some anomalous TTVs, that do not fall on the analytically expressed peaks -- more significantly the closer you get in orbital period ratio space. These are likely other aliases of first-order super-periods. However, as the majority of TTV signals fall along our analytic solutions (see Figure~\ref{fig: analytic numerical landscape comp} \& later discussion), we can use these to define efficient priors for modeling TTVs from perturbing planets.

From an analytic standing, we now have a sense of where the multi-modalities are likely to arise in period ratio space. One could imagine that when a TTV is observed, in characterizing the perturbing planets orbital characteristics, we essentially would start by drawing a horizontal line across Figure~\ref{fig: analytic numerical landscape} and then any intersection would be a possible period ratio solution. In practice, this would be done typically with a numerical simulator and some type of sampling model (e.g., MCMC or nested sampling). However, as previously stated, adopting a diffuse prior on perturber period, will result in a posterior that tends to get stuck in one of the orbital period modes.

One would thus want to define uniform priors that span from the peak in TTV period space to the valley where another super-period takes over as the dominant effect. That is, to test orbital period ratios ranging from 1/10 to 10, one should adopt uniform priors spanning the orbital period ratio ranges corresponding to the first-order super-periods, as shown in Table~\ref{table: priors}

\begin{table}[h!] 
    \footnotesize
    \centering
    \begin{tabular}{|c|c|c|}
        \hline
        Prior Name & Period Ratio Range &  $j:k$ Super-Period  \\  \hline
        $\pi_{-21}$ & $\mathcal{U}\left[\frac{1}{10}, \, \frac{2}{19}\right]$ & $2:1$, $m=-8$ \\ \hline
        $\pi_{-20}$ & $\mathcal{U}\left[\frac{2}{19}, \, \frac{1}{9}\right]$ &  $2:1$, $m=-7$ \\ \hline
        $\pi_{-19}$ & $\mathcal{U}\left[\frac{1}{9}, \, \frac{2}{17}\right]$ & $2:1$, $m=-7$  \\ \hline
        $\pi_{-18}$ & $\mathcal{U}\left[\frac{2}{17}, \, \frac{1}{8}\right]$ & $2:1$, $m=-6$  \\ \hline
        $\pi_{-17}$ & $\mathcal{U}\left[\frac{1}{8}, \, \frac{2}{15}\right]$ & $2:1$, $m=-6$  \\ \hline
        $\pi_{-16}$ & $\mathcal{U}\left[\frac{2}{15}, \, \frac{1}{7}\right]$ & $2:1$, $m=-5$  \\ \hline
        $\pi_{-15}$ & $\mathcal{U}\left[\frac{1}{7}, \, \frac{2}{13}\right]$ & $2:1$, $m=-5$  \\ \hline
        $\pi_{-14}$ & $\mathcal{U}\left[\frac{2}{13}, \, \frac{1}{6}\right]$ & $2:1$, $m=-4$ \\ \hline
        $\pi_{-13}$ & $\mathcal{U}\left[\frac{1}{6}, \, \frac{2}{11}\right]$ &  $2:1$, $m=-4$ \\ \hline
        $\pi_{-12}$ & $\mathcal{U}\left[\frac{2}{11}, \, \frac{1}{5}\right]$ & $2:1$, $m=-3$ \\ \hline
        $\pi_{-11}$ & $\mathcal{U}\left[\frac{1}{5}, \, \frac{2}{9}\right]$ &  $2:1$, $m=-3$ \\ \hline
        $\pi_{-10}$ & $\mathcal{U}\left[\frac{2}{9}, \, \frac{1}{4}\right]$ & $2:1$, $m=-2$ \\ \hline
        $\pi_{-9}$ & $\mathcal{U}\left[\frac{1}{4}, \, \frac{2}{7}\right]$ & $2:1$, $m=-2$ \\ \hline
        $\pi_{-8}$ & $\mathcal{U}\left[\frac{2}{7}, \, \frac{1}{3}\right]$ & $2:1$, $m=-1$ \\ \hline
        $\pi_{-7}$ & $\mathcal{U}\left[\frac{1}{3}, \, \frac{3}{8}\right]$ & $2:1$, $m=-1$ \\ \hline
        $\pi_{-6}$ & $\mathcal{U}\left[\frac{3}{8}, \, \frac{2}{5}\right]$ & $3:2$, $m=-2$ \\ \hline
        $\pi_{-5}$ & $\mathcal{U}\left[\frac{2}{5}, \, \frac{3}{7}\right]$ & $3:2$, $m=-2$ \\ \hline
        $\pi_{-4}$ & $\mathcal{U}\left[\frac{3}{7}, \, \frac{1}{2}\right]$ & $2:1$ \\ \hline
        $\pi_{-3}$ & $\mathcal{U}\left[\frac{1}{2}, \, \frac{3}{5}\right]$ & $2:1$ \\ \hline
        $\pi_{-2}$ & $\mathcal{U}\left[\frac{3}{5}, \, \frac{2}{3}\right]$ & $3:2$ \\ \hline
        $\pi_{-1}$ & $\mathcal{U}\left[\frac{2}{3}, \, 1\right]$ & $3:2$ \\ \hline
        $\pi_1$ & $\mathcal{U}\left[1, \, \frac{3}{2}\right]$ & $2:3$ \\ \hline
        $\pi_2$ & $\mathcal{U}\left[\frac{3}{2}, \, \frac{5}{3}\right]$ & $2:3$ \\ \hline
        $\pi_3$ & $\mathcal{U}\left[\frac{5}{3}, \, 2\right]$ & $1:2$ \\ \hline
        $\pi_4$ & $\mathcal{U}\left[2, \, \frac{7}{3}\right]$ & $1:2$ \\ \hline
        $\pi_5$ & $\mathcal{U}\left[\frac{7}{3}, \, \frac{5}{2}\right]$ & $2:3$, $m=-2$ \\ \hline
        $\pi_6$ & $\mathcal{U}\left[\frac{5}{2}, \, \frac{8}{3}\right]$ & $2:3$, $m=-2$ \\ \hline
        $\pi_7$ & $\mathcal{U}\left[\frac{8}{3}, \, 3\right]$ & $2:3$, $m=-1$ \\ \hline
        $\pi_8$ & $\mathcal{U}\left[3, \, \frac{7}{2}\right]$ & $2:3$, $m=-1$  \\ \hline
        $\pi_9$ & $\mathcal{U}\left[\frac{7}{2}, \, 4\right]$ & $3:4$, $m=-2$  \\ \hline
        $\pi_{10}$ & $\mathcal{U}\left[4, \, \frac{9}{2}\right]$ & $3:4$, $m=-2$  \\ \hline
        $\pi_{11}$ & $\mathcal{U}\left[\frac{9}{2}, \, 5\right]$ & $4:5$, $m=-3$ \\ \hline
        $\pi_{12}$ & $\mathcal{U}\left[5, \, \frac{11}{2}\right]$ & $4:5$, $m=-3$  \\ \hline
        $\pi_{13}$ & $\mathcal{U}\left[\frac{11}{2}, 6\right]$ &  $5:6$, $m=-4$\\ \hline
        $\pi_{14}$ & $\mathcal{U}\left[6, \, \frac{13}{2}\right]$ & $5:6$, $m=-4$ \\ \hline
        $\pi_{15}$ & $\mathcal{U}\left[\frac{13}{2}, 7\right]$ & $6:7$, $m=-5$  \\ \hline
        $\pi_{16}$ & $\mathcal{U}\left[7, \, \frac{15}{2}\right]$ & $6:7$, $m=-5$  \\ \hline
        $\pi_{17}$ & $\mathcal{U}\left[\frac{15}{2}, 8\right]$ & $7:8$, $m=-6$ \\ \hline
        $\pi_{18}$ & $\mathcal{U}\left[8, \, \frac{17}{2}\right]$ & $7:8$, $m=-6$ \\ \hline
        $\pi_{19}$ & $\mathcal{U}\left[\frac{17}{2}, 9\right]$ & $8:9$, $m=-7$ \\ \hline
        $\pi_{20}$ & $\mathcal{U}\left[9, \, \frac{19}{2}\right]$ & $8:9$, $m=-7$ \\ \hline
        $\pi_{21}$ & $\mathcal{U}\left[\frac{19}{2}, 10\right]$ & $9:10$, $m=-8$ \\ \hline
    \end{tabular}
    \caption{Prior names, orbital period range that each prior spans, and the corresponding super-period that analytically explains this mode of the unseen perturbing planet's period. Here, $\pi_i =  \mathrm{P}(\bm{\theta}|\mathcal{M}_i)$, the local prior on free parameters. }
    \label{table: priors}
\end{table}

These period ranges can also be seen in Figure~\ref{fig: analytic numerical landscape}. By sampling the TTV inversion problem in these period ranges, we can avoid the multi-modality otherwise expected in single-planet TTVs, in each period range there should only one solution. Given this set of prior ranges, motivated by both numerical and analytic solutions, one could then (i) calculate the model evidence of a perturbing planet solution in a single prior range and (ii) correctly sample the full posterior distribution of perturbing planet properties, combining each individual posterior using Bayesian model averaging and using the individual evidences as weights.

\subsection{Agreement Between Analytic and Numerical Predictions}

\begin{figure*}[ht!]
    \centering
    \includegraphics[width=\textwidth]{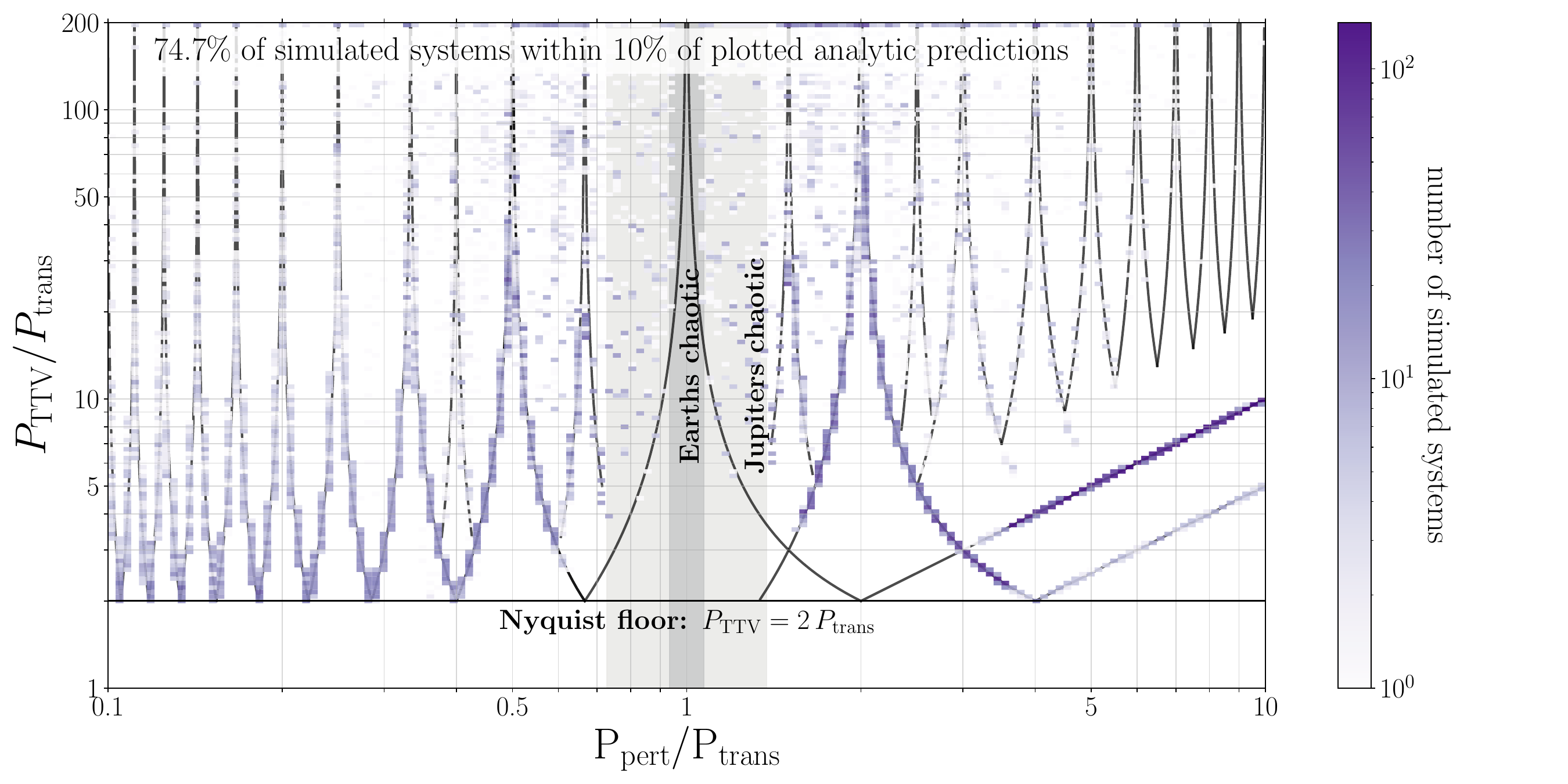}
    \caption{
        Two-dimensional histogram of numerical simulations and analytic predictions. Black lines show the analytic TTV predictions. 74.7\% of simulated systems fall within 10\% of analytic predictions from the TTV circus tent diagram.
    }
    \label{fig: analytic numerical landscape comp}
\end{figure*}

We investigated the degree of agreement between the numerical simulations and the analytic predictions for the dominant TTV period. To perform this comparison, we considered all numerical simulations that passed the stability criteria and the amplitude selection criteria, and evaluated whether the recovered maximum TTV period from the simulation is consistent with the analytic prediction to within 10\%. We find that in 74.7\% of cases, the simulated TTV periods agree with the analytic predictions at this level of precision. This is shown in Figure~\ref{fig: analytic numerical landscape comp} as a two-dimensional histogram of the number of numerical simulations in specific regions of TTV period vs. orbital period ratio space in the circus tent diagram.

As can be seen in Figure~\ref{fig: analytic numerical landscape comp}, the fraction of simulations that are inconsistent with the analytic predictions increases as the period ratio approaches unity. Some of these discrepancies likely reflect the presence of additional first-order near-resonance super-periods not explicitly included in our analytic framework. When restricting the analysis to systems with period ratios either below 0.5 or above 2, we find that the fraction of simulations consistent with the analytic predictions increases to 83.1\%.

Additionally, as can be seen in Figure~\ref{fig: orbital_landscape_combined_amplitude}, many of the systems that do not fall on the analytically predicted curves have very large TTV amplitudes, suggestive of a significant dynamic interaction that is not likely to be caused by a stable planet-planet interaction. The largest observed planet-planet TTV to date is $\sim$2 days from TOI-4504\,c \citep{Vitkova2025}. Before that, the largest known planet-planet TTV was KOI-142\,b with a $\sim$12 hour TTV \citep{Nesvorny2013}. If we remove all TTVs with an amplitude larger than $10^3$ minutes we find that the fraction of simulations consistent with the analytic predictions again increases, this time to 84.3\%.

Even in cases where the analytic super-period predictions fail to match the simulated TTV periods precisely, we emphasize that our proposed fitting framework via period prior binning remains effective. In practice, modelers fitting observed TTVs would employ a full dynamical integrator such as \texttt{TTVFast} to evaluate candidate solutions within these discrete period-ratio prior bins. Using a robust dynamical model in each specific period prior bin, the correct solution can still be recovered, even when the simple analytic super-period estimate deviates from the actual TTV period.

Lastly, we note that although the analytic predictions for the TTV period are strictly independent of planetary mass, our simulations exhibit a small residual dependence on the masses and mass ratios of the planet pairs. This likely arises from two effects. First, the analytic expressions track only some of the dominant super-periods and the synodic period, whereas in certain regions of mass–eccentricity space these may not be necessarily the highest-amplitude components of the full TTV signal. Second, the TTV periods extracted from the numerical simulations are obtained by fitting a single-sinusoid LS periodogram to the simulated transit times; for systems with multi-periodic or non-sinusoidal structure, this procedure can identify an alias or a blended frequency rather than the true dominant mode. Together, these effects introduce minor mass-dependent deviations from the idealized analytic curves.

\subsection{The Exoplanet Edge} \label{sec: edge}

There is one anomalous effect besides the otherwise super-period ``peak'' dominated orbital landscape: the two ``edges'' found at orbital period ratios $\geq$ 4. As shown in Figure~\ref{fig: analytic numerical landscape}, the TTV periods of these two edges are equal to the period of the perturbing planet and half the period of the perturbing planet. Interestingly, we find that TTVs are not observed with periods faster than half the orbital period of the perturbing planet.

Thus, in addition to shedding light in the multi-modality of low-eccentricity single-planet systems with TTVs, these analytic and numerical simulations reveal another effect -- a minimum recoverable TTV for external perturbers with periods. This limiting ``exoplanet edge'' for planet-planet TTVs is discussed in depth in \citet{Yahalomi2025b}, where it is demonstrated that the exoplanet edge persists for high eccentricity systems.

\subsection{Testing this Modeling Framework}

The goal of this study is to present the theoretical motivation for fitting planet-planet TTVs in single-transiting systems by breaking the parameter space of possible perturbing periods into discrete prior bins, thereby naturally accounting for multi-modalities in perturbing period space. We have motivated this proposed period prior bin structure through both analytic theory and numerical simulations.

The morphology of TTV signals is notoriously complex, depending sensitively on both the details of the planetary system architecture (i.e., planetary dynamics) and observational constraints of transit observations. In this work, we chose not to include a specific case study in which an observed system is analyzed using the proposed framework. While such an example could demonstrate the method on a well-characterized system (e.g., artificially withholding knowledge of the perturbing planet period in a well-classified planet-planet TTV system), it would not constitute a comprehensive assessment of the framework's robustness and thus may in fact be misleading. A full evaluation of the method’s performance will require testing across a diverse set of dynamical architectures, mass ratios, orbital configurations, and observational constraints that jointly shape the observable TTV signal. Such a study is beyond the scope of this paper, but represents a natural direction for future work aimed at validating this approach under realistic observational conditions. We would recommend any such study to at least include both comprehensively studied near-resonant planet-planet TTV systems with sinusoidal TTVs (e.g., Kepler-19 \citep{Ballard2011, Malavolta2017}) as well as systems with more exotic -- that is less sinusoidal -- TTV shapes (e.g., Kepler-46 / KOI-872 \citep{Nesvorny2012}.

\section{Parameter Estimation and Model Selection}
\label{sec: model comparison}

We now turn to the problem of how to fit the TTVs of a transiting planet (with no other known companions) that exhibits TTVs. First, we highlight that simply throwing an MCMC or nested sampling algorithm at the data with very broad priors is problematic. The modes are too numerous (infinite in fact), too densely packed and too sharp for algorithms to reliably recover them. If modes are missed, not only are the posteriors inaccurate but so too is the model evidence - which is crucial if one wishes to rank competing models (e.g. an exomoon). These problems are so extreme that they have largely prevented the community from inferring the landscape of allowed perturbing planet solutions to date.

As an aside, one might question the point of even attempting this. After all, even a correctly sampled posterior will inevitably be highly multi-modal still. It is unlikely we will arrive at a unique solution, except in rare cases like KOI-872 which exhibit outstanding signal-to-noise \citep{Nesvorny2012}. But a multi-modal posterior is still useful, constraining the global architecture of the system, especially when combined with other measurements such as radial velocities or astrometry. Further, the Bayesian evidence can at last be correctly estimated, allowing one to weigh different models. But perhaps the most important argument is this - there are thousands of such systems and we will likely find many, many more in the years and decades ahead. Each example is a signal, a whisper of that particular system's configuration. Combined together, they become a chorus speaking to us about the demographics of exoplanets more broadly. While the existence of unexplained TTVs already suggests the likely presence of perturbing planets, identifying the discrete set of candidate orbital periods under physically motivated assumptions significantly sharpens our understanding of these systems. This allows us to place stronger constraints on the architectures of planetary systems, inform population-level statistics, and guide targeted follow-up efforts aimed at resolving these degeneracies through complementary observations.

\subsection{Sampling the Global Posterior}

The key to making progress is to leverage the map revealed in this work. With the $N$ appropriate period priors now established, one could then run $N$ fits in each window. In each fit, one would follow the standard process regressing a TTV model to the data, but the prior on the perturbing planet's period would be bounded by one of the perturbing period ranges commensurable with the first-order super-period modes. The global posterior is then found by combining these local solutions via Bayesian model averaging.

Consider that from each local fit, we have a unique prior on the free parameters, $\bm{\theta}$, defined as $\mathrm{P}(\bm{\theta}|\mathcal{M}_i)$, where $\mathcal{M}_i$ denotes the conditional of adopting the model that uses the $i^{\mathrm{th}}$ window. Further, we obtain a local posterior distribution on $\bm{\theta}$ from each fit defined as $\mathrm{P}(\bm{\theta}|\mathcal{D},\mathcal{M}_i)$, where $\mathcal{D}$ represents the transit timing data (common to all models). The local posterior can be defined via Bayes' theorem as:

% annotated version
% annotated version
\begin{equation} \label{eq:local bayes}
\underbrace{\mathrm{P}(\bm{\theta}|\mathcal{D}, \mathcal{M}_i)}_{\shortstack{\scriptsize $\mathrm{local}\,\,\bm{\theta}$\\ \scriptsize $\mathrm{posterior}$}} = \frac{ \overbrace{\mathrm{P}(\mathcal{D}|\bm{\theta}, \mathcal{M}_i)}^{\shortstack{\scriptsize $\mathrm{local}\,\,\bm{\theta}$ \\ \scriptsize $\mathrm{evidence}$}} \overbrace{\mathrm{P}(\bm{\theta}|\mathcal{M}_i)}^{\shortstack{\scriptsize $\mathrm{local}\,\,\bm{\theta}$ \\ \scriptsize $\mathrm{prior}$}} }{ \underbrace{\mathrm{P}(\mathcal{D}|\mathcal{M}_i)}_{\shortstack{\scriptsize $\mathrm{local\,\,model}$ \\ \scriptsize $\mathrm{evidence}$}} }.
\end{equation}

The global posterior can now be found by performing a weighted sum of each posterior, with the weights set by the probability of each model, given the data:

% simple version
%\begin{align}
%\mathrm{P}(\bm{\theta}|\mathcal{D}) &= \sum_{i=1}^N \mathrm{P}(\bm{\theta}|\mathcal{D},\mathcal{M}_i) \mathrm{P}(\mathcal{M}_i|\mathcal{D}) 
%\end{align}

% annotated version
\begin{equation}
\underbrace{\mathrm{P}(\bm{\theta}|\mathcal{D})}_{\shortstack{\scriptsize $\mathrm{global}\,\,\bm{\theta}$ \\ \scriptsize $\mathrm{posterior}$}} = \displaystyle \sum_{i=1}^N \underbrace{\mathrm{P}(\bm{\theta}|\mathcal{D},\mathcal{M}_i)}_{\shortstack{\scriptsize $\mathrm{local}\,\,\bm{\theta}$ \\ \scriptsize $\mathrm{posterior}$}} \underbrace{\mathrm{P}(\mathcal{M}_i|\mathcal{D})}_{\shortstack{\scriptsize $\mathrm{local\,\,model}$ \\ \scriptsize $\mathrm{posterior}$}}.
\end{equation}

In the above, the weights, $\mathrm{P}(\mathcal{M}_i|\mathcal{D})$, should not be confused with the Bayesian evidences, $\mathrm{P}(\mathcal{D}|\mathcal{M}_i)$. However they are intimately related via Bayes' theorem as

% simple version
%\begin{align}
%\mathrm{P}(\mathcal{M}_i|\mathcal{D}) &= \frac{ \mathrm{P}(\mathcal{D}|\mathcal{M}_i) \mathrm{P}(\mathcal{M}_i) }{ \mathrm{P}(\mathcal{D}) }
%\end{align}

% annotated version
\begin{equation}
\underbrace{\mathrm{P}(\mathcal{M}_i|\mathcal{D})}_{\shortstack{\scriptsize $\mathrm{local\,\,model}$ \\ \scriptsize $\mathrm{posterior}$}} = \frac{ \overbrace{\mathrm{P}(\mathcal{D}|\mathcal{M}_i)}^{\shortstack{\scriptsize $\mathrm{local\,\,model}$ \\ \scriptsize $\mathrm{evidence}$}} \overbrace{\mathrm{P}(\mathcal{M}_i)}^{\shortstack{\scriptsize $\mathrm{local\,\,model}$ \\ \scriptsize $\mathrm{prior}$}} }{ \underbrace{\mathrm{P}(\mathcal{D})}_{\shortstack{\scriptsize $\mathrm{global\,\,model}$ \\ \scriptsize $\mathrm{evidence}$}} }.
\end{equation}

The above is useful since it allows one to inject demographics priors into the inference. For example, based on previous exoplanet demographics studies, one might wish to down-weight certain windows where planets are rare, which can be achieved using $\mathrm{P}(\mathcal{M}_i)$. Finally, to complete the procedure, we need to evaluate $\mathrm{P}(\mathcal{D})$. Since this term acts like a normalization constant, it can be found by the demand for all probabilities to sum to unity, such that

% simple version
%\begin{align}
%\mathrm{P}(\mathcal{D}) &= \sum_{j=1}^N \mathrm{P}(\mathcal{D}|M_j) \mathrm{P}(M_j) 
%\label{eqn:globevidence}
%\end{align}

% annotated version
\begin{equation}
\underbrace{\mathrm{P}(\mathcal{D})}_{\shortstack{\scriptsize $\mathrm{global\,\,model}$ \\ \scriptsize $\mathrm{evidence}$}} = \displaystyle \sum_{j=1}^N \underbrace{\mathrm{P}(\mathcal{D}|\mathcal{M}_j)}_{\shortstack{\scriptsize $\mathrm{local\,\,model}$ \\ \scriptsize $\mathrm{evidence}$}} \underbrace{\mathrm{P}(\mathcal{M}_j)}_{\shortstack{\scriptsize $\mathrm{local\,\,model}$ \\ \scriptsize $\mathrm{prior}$}}.
\label{eqn:globevidence}
\end{equation}

As a final note, one might be concerned that our defined windows do not guarantee a uni-modal solution, since for example one of our assumptions is low-eccentricity. Accordingly, the use of a multi-modal fitting algorithm is still advised, to account for this possibility. Nevertheless, the number of modes will be far more manageable in these bite-size windows than the entire parameter volume and should lead to more reliable results.

% OLD TEXT...
%In order to compare the $N$ resulting models with the different period priors, $\alpha_\mathrm{i}$, one could then compare the Bayesian evidence, $Z_\mathrm{i}$, of each of these $N$ models. Bayesian evidence for a given model, M$_\mathrm{i}$, is defined as
%
%\begin{equation} \label{eq: bayes_evidence}
%    Z_\mathrm{i} = P(\mathcal{D}|  \theta, M_\mathrm{i}, \alpha_\mathrm{i}) .
%\end{equation}
%
%This model with the highest Bayesian evidence would be selected as the best fit planet-planet model. Additionally, one could sum the complete set of possible perturbing period modes via Bayesian model averaging with the individual evidences used as weights, in order to determine the Bayesian evidence for the perturbing planet spanning all of the different modes.

\subsection{Planets vs. Moons}

At this point, it's fairly straight-forward to extend our results to compare the assumed perturbing planet model to some other model, for example an exomoon as being responsible for the TTVs. It should be noted in the last subsection, all of the probability distributions are implicitly conditional upon the global model - a perturbing planet. Thus, really one should adjust $\mathrm{P}(\mathcal{D}) \to \mathrm{P}(\mathcal{D}|\mathrm{PP})$ where ``PP'' denotes perturbing planet. Thus, Equation~\ref{eqn:globevidence} provides the overall evidence of the PP model.

The user now need only execute an additional fit for the moon model(s), which we dub ``PM''. The odds ratio between the two models will then be given as

\begin{equation}
\frac{ \mathrm{P}(\mathrm{PP}|\mathcal{D}) }{ \mathrm{P}(\mathrm{PM}|\mathcal{D}) } = \frac{ \mathrm{P}(\mathcal{D}|\mathrm{PP}) }{ \mathrm{P}(\mathcal{D}|\mathrm{PM}) } \frac{ \mathrm{P}(\mathrm{PP}) }{ \mathrm{P}(\mathrm{PM}) }.
\end{equation}

The ratio of the model priors is particularly challenging to estimate given the dearth of exomoon detections, and thus practically speaking setting this ratio to unity - equivalent to a Bayes factor - would be a possible path forward.

\section{Conclusion}

Transit timing variations (TTVs) are found in many transiting planet systems. When uncovered in (ostensibly) single-planet systems, the TTVs induced by a perturbing (non-transiting) planet are difficult to characterize -- as the solution is highly multi-modal with respect to the unseen planet's orbital period.

As multi-planet systems have been found to date to generally have low eccentricities, we adopt a pragmatic approach and study the orbital landscape of nearly circular planet-planet TTVs. This eccentricity space has the advantage of being dominated by first-order super-periods, allowing for us to analytically map out the location of these multi-modalities in orbital period ratio space.

Analytic formulae of first-order super-periods and their aliases, as well as numerical ($N$-body) simulations reveal that there are quantifiable modes in orbital period ratio space commensurable with a given TTV period. Therefore, in order to efficiently and accurately sample the complete parameter space of possible perturbing planets responsible for TTVs observed in single-planet systems, one must adopt priors informed by these super-periods and their aliases. We provide a set of orbital period ratios for which period priors should be defined. We then demonstrate how one could use these priors in order to effectively sample the complete parameter space for perturbing planets, and then adopt a Bayesian model comparison approach to determine whether a TTV signal is more likely cause by an unseen perturbing planet or an unseen perturbing moon. We encourage future work to test this model comparison approach via injection recovery of planet-planet systems and planet-moon systems.

Additionally, we uncover that perturbing planets don't induce observable TTVs with a dominant period faster than half their own orbital period. As a result, TTVs observed with a dominant period faster than this limiting ``exoplanet edge'' are suggestive of additional mass in the system -- thus providing a new path to uncovering new exoplanets and perhaps exomoons. For more on this, see \citet{Yahalomi2025b}.

\section{Acknowledgments}

We thank the anonymous reviewer who's feedback improved this manuscript. The authors are deeply grateful to Eric Agol, David Nesvorny, Daniel Fabrycky and Matthew Holman for inspiring conversations.

D.A.Y. and D.K. acknowledge support from NASA Grant \#80NSSC21K0960.

D.A.Y. acknowledges support from
the NASA/NY Space Grant

D.A.Y. thanks the LSST-DA Data Science Fellowship Program, which is funded by LSST-DA, the Brinson Foundation, and the Moore Foundation; his participation in the program has benefited this work.

D.A.Y and D.K. thank the following for their generous support to the Cool Worlds Lab:
Douglas Daughaday,
Elena West,
Tristan Zajonc,
Alex de Vaal,
Mark Elliott,
Stephen Lee,
Zachary Danielson,
Chad Souter,
Marcus Gillette,
Tina Jeffcoat,
Jason Rockett,
Tom Donkin,
Andrew Schoen,
Reza Ramezankhani,
Steven Marks,
Nicholas Gebben,
Mike Hedlund,
Leigh Deacon,
Ryan Provost,
Nicholas De Haan,
Emerson Garland,
The Queen Road Foundation Inc,
Scott Thayer,
Frank Blood,
Ieuan Williams,
Xinyu Yao,
Axel Nimmerjahn,
Brian Cartmell,
\&
Guillaume Le Saint.



\subsection{Software}
\texttt{matplotlib} \citep{matplotlib}, $\,$
\texttt{numpy} \citep{numpy}, $\,$
\texttt{scipy} \citep{scipy}, $\,$
\texttt{TTVFast} \citep{Deck2014}, $\,$ \texttt{ChatGPT} was
utilized to improve wording at the sentence level and assist with
coding inquires -- last accessed in 2025 June.



\newpage
\bibliography{main}{}
\bibliographystyle{aasjournal}




\end{document}